\documentclass[conference, 11pt]{IEEEtran}
\IEEEoverridecommandlockouts
\usepackage{algorithm}
\usepackage[noend]{algpseudocode}
\usepackage{cite}
\usepackage{amsmath,amssymb,amsfonts}
\usepackage{graphicx}
\usepackage{textcomp}
\usepackage{xcolor}
\usepackage{braket}
\usepackage{hyperref}
\usepackage{fancyvrb}

\algblock{Inputs}{EndInputs}
\algnotext{EndInputs}
\algblock{Outputs}{EndOutputs}
\algnotext{EndOutputs}

\def\BibTeX{{\rm B\kern-.05em{\sc i\kern-.025em b}\kern-.08em
    T\kern-.1667em\lower.7ex\hbox{E}\kern-.125emX}}

\newcommand{\aref}[1]{\hyperref[#1]{Appendix~\ref*{#1}}}
\begin{document}

\title{A Q\# Implementation of a Quantum Lookup Table for Quantum Arithmetic Functions}

\author{\IEEEauthorblockN{Rajiv Krishnakumar}
\IEEEauthorblockA{\textit{Core R\&D} \\
\textit{Goldman Sachs}\\
Geneva, Switzerland \\
rajiv.krishnakumar@gs.com}
\and
\IEEEauthorblockN{Mathias Soeken}
\IEEEauthorblockA{\textit{Microsoft Quantum} \\
\textit{Microsoft}\\
Z{\"u}rich, Switzerland \\
mathias.soeken@microsoft.com}
\and
\IEEEauthorblockN{Martin Roetteler}
\IEEEauthorblockA{\textit{Microsoft Quantum} \\
\textit{Microsoft}\\
Redmond, WA, USA \\
martin.roetteler@microsoft.com}
\and
\IEEEauthorblockN{William Zeng}
\IEEEauthorblockA{\textit{Core R\&D} \\
\textit{Goldman Sachs}\\
New York, NY, USA \\
william.zeng@gs.com}
}

\maketitle

\begin{abstract}
In this paper, we present Q\# implementations for arbitrary single-variabled fixed-point arithmetic operations for a gate-based quantum computer based on lookup tables (LUTs). 
In general, this is an inefficent way of implementing a function since the number of inputs can be large or even infinite. 
However, if the input domain can be bounded and there can be some error tolerance in the output (both of which are often the case in practical use-cases), 
the quantum LUT implementation of certain quantum arithmetic functions can be more efficient than their corresponding reversible arithmetic implementations. 
We discuss the implementation of the LUT using Q\# and its approximation errors. We then show examples of how to use the LUT to implement quantum arithmetic functions and 
compare the resources required for the implementation with the current state-of-the-art bespoke implementations of some commonly used arithmetic functions.
The implementation of the LUT is designed for use by practitioners to use when implementing end-to-end quantum algorithms. In addition, 
given its well-defined approximation errors, the LUT implementation makes for a clear
 benchmark for evaluating the efficiency of bespoke quantum arithmetic circuits .
\end{abstract}


\section{Introduction}
The purpose of a mathematical function (within a piece of code or even in a general sense) is to map an input $x$ to an output $f(x)$ \cite{al-tusi, euler}.
When implementing this on a computer, the most general method is to create a lookup table (LUT)---a structure
that stores all of the possible inputs along with their corresponding outputs. In this paper, we present a Q\#~\cite{qsharp} implementation of such a 
structure for a gate-based quantum computer for fixed-point arithmetic functions.\footnote{https://github.com/microsoft/QuantumLibraries/pull/611} 
We focus on single variable real functions i.e. $f: \mathbb{R} \rightarrow \mathbb{R}$.  In essence, 
the implemented LUT executes the unitary $U : \ket{x}\ket{00\dots 0} \mapsto \ket{x}\ket{f(x)}$ where $x$ and $f(x)$ are 
fixed-point representations of the input and output values. At first this may seem like an inefficent way of implementing a function 
since the number of inputs can be large or even infinite. However, if the input domain can be bounded and there can be 
some error tolerance in the output (both of which are often the case in practical use-cases), 
the quantum LUT implementation of certain quantum arithmetic functions can be more efficient than their corresponding 
reversible arithmetic implementations. In the rest of this paper, we will first discuss the implementation of the LUT in
Q\# using the convenient built-in functions in Q\# that make the implemention well-structured, followed by how the implementation is geared for practitioners that are interested in implementing 
quantum arithmetic operations within a larger end-to-end algorithm. We then discuss the
approximation error of the operator, which is a key factor in enabling this implementation to be used in end-to-end algorithms
as well as a benchmark for evaluating the efficiency of single variable bespoke quantum arithmetic circuits. We will then show examples of
how to use the LUT to implement specific quantum arithmetic functions and compare the resources required for the
implementation with the current state-of-the-art bespoke implementations of the exponential, Gaussian and square root functions.

\subsection*{Related work}
While we compare our method to some dedicated arithmetic implementations in the
remainder, we want to point out some related general purpose work and how it
differentiates from our work.  In~\cite{HRS18}, the authors proposed a method to
implement single-variabled fixed-point arithmetic using piecewise polynomial
evaluation.  In order to automate their approach the Remez
algorithm~\cite{Remez} can be used to determine a piecewise polynomial
approximation given as input an $L_\infty$-error on the respective domain.  The
precision of the fixed-point numbers rely on the coefficients in the polynomial
approximation and the implementation of arithmetic operations to evalaute the
polynomial.  The approach works well for functions in which the function values
of the function in the evaluated input domain are contained in a small interval.

In~\cite{LHRS}, the authors presented an algorithm that creates a quantum
implementation based on a classical Boolean function, which can be provided
symbolically in terms of a classical logic network.  Such and similar so-called
hierarchical reversible synthesis methods can be used given as input an
approximation of a single-variabled fixed-point function.  For these algorithms,
the effort is shifted towards classical optimization of the logic network for
which existing automatic optimization algorithms can be leveraged.

\section{Implementation and of a Quantum Lookup Table Circuit in Q\#}
In this section, we will first discuss the the implementation of a LUT in Q\# for arithmetic functions. In \autoref{subsec:FxPLUT} we introduce
the fixed-point \textsc{SelectSwap} network which is at the crux of the implementation, and discuss its implementation in Q\#. Then, in  
\autoref{subsec:LUTforArith} we will discuss how the function that implements this quantum circuit 
(such functions are referred to as \textit{operations} in Q\#) can be used within a wrapper
function to allow the user to easily create a quantum LUT operator for any desired arithmetic function.

\subsection{The fixed-point representation implementation of a LUT}
\label{subsec:FxPLUT}

\begin{figure*}[htbp]
\centerline{\includegraphics[width=0.8\textwidth]{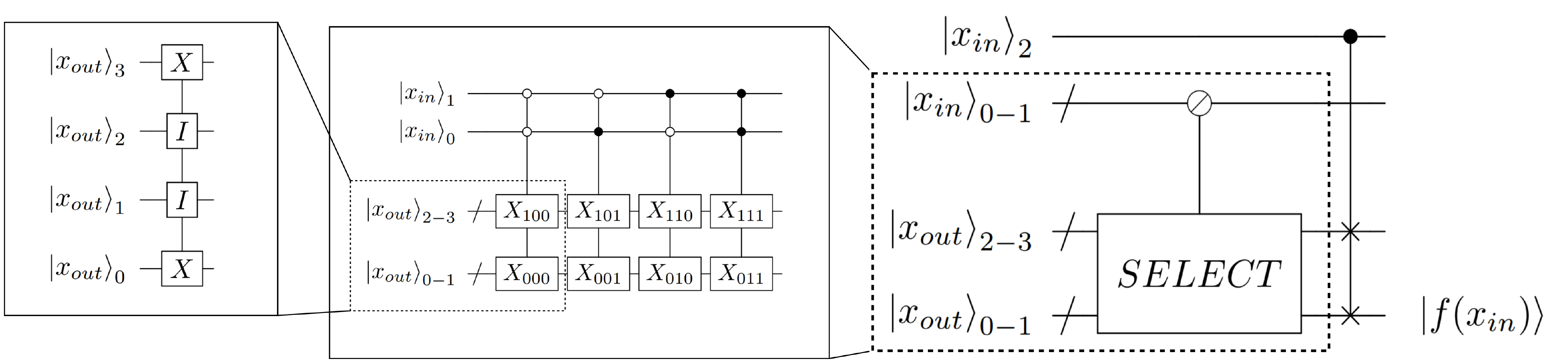}}
\caption{Example of a \textsc{SelectSwap} network with a 3-qubit input and 2-qubit output.}
\label{fig:simpleselectswap}
\end{figure*}

To implement the generic fixed-point representation LUT, we use the \textsc{SelectSwap} network \cite{LKS18}, for which we present a
running example in \autoref{fig:simpleselectswap}. To understand the network, we can start by looking at the standalone \textsc{Select} network~\cite{BGB+18}.
This network is a series of $X$ gates on the output qubits controlled on the input qubits. For example, with a two qubit input
of $\ket{00}$ and a desired two qubit output of $\ket{10}$, the first layer of the \textsc{Select} network would be
an $X$ gate on the first output qubit controlled on all 3 input qubits being $0$. The next layer of the network is constructed
using the desired output for the input $\ket{01}$ and so on.
In general, this requires a T-depth of $2^{n}$ where $n$ is the number input qubits.
However we can create a trade-off between T-depth and width of the circuit by adding a $\textsc{Swap}$ network. 
To do this, we first
compute several outputs in parallel based on the least significant input qubits, and then swap in the desired output based on the
most significant input qubits. Building upon on the previous example, let us look at the case where we have a three qubit input and
two qubit output, with the input $\ket{000}$ having a desired output of $\ket{10}$ and the input $\ket{100}$ having a desired output of $\ket{01}$.
We proceed by selecting on the 2 least significant qubits and then using the most significant qubit as the \textit{swap qubit}. 
To implement this, we first construct a 4 qubit output \textsc{Select} network which would output $\ket{10}\ket{01}$ (constructed using controlled-$X$ gates) 
conditioned the last two input qubits being $0$. The last step is to ensure that the first 2 qubits of our output state ends up begin the 
register containing $f(x)$. In this example, if the first input qubit is also $\ket{0}$, then we don't have to do anything. But if it is $\ket{1}$, 
then we would like to swap the output registers. So to implement the first layer of the $\textsc{Swap}$ network, 
we swap the last two output qubits with the first two output qubits controlled on the
most significant input qubit being $1$.

In general, the \textsc{SelectSwap} network requires a T-depth of
\begin{align} 
	2^{n-l}+l
	\label{eqn:t-depth}
\end{align}
and a qubit count of 
\begin{align}
	m \times 2^{l}
	\label{eqn:qubit-count}
\end{align}
where $n$ is the number input qubits, $l$ is the number of swap qubits and $m$ is
number of qubits required to store the output $f(x)$ in a fixed-point register.

So far we have described our \textsc{SelectSwap} network using inputs and outputs of binary qubit registers. However we would like
to abstract the binary qubit regsiter to a fixed-point qubit register. Q\# has an in-built fixed point register structure which readily takes care of the
conversions between binary and fixed-point qubit registers. That in addition to Q\# having some readily available operations in its in-built libraries (e.g. \textit{MultiplexOperations})
 allowed us to implement the fixed-point  \textsc{SelectSwap} operation in fewer than 100 lines of code. 

\subsection{Using the fixed-point implementation of a LUT to implement quantum arithmetic functions}
\label{subsec:LUTforArith}

The \textsc{SelectSwap} operation from the previous section requires the user to input the number of qubits in the input and output fixed-point
register, as well as the binary values of each of the desired outputs. However, to remove the burden 
of computing these quantities from the user, we created a wrapper function \textsc{ApplyFunctionWithLookup} that 
contains the above \textsc{SelectSwap} network. This wrapper function requires the user to provide the input domain $(x_{\min}, x_{\max})$, maximum allowed input error tolerance $\epsilon_{\mathrm{in}}$,
minimum required output precision $\epsilon_{\mathrm{out}}$ and number of swap qubits. Aside from the last one, these are the parameters that practitioners will typically focus on when implementing
quantum arithmetic operations within a larger end-to-end algorithm. The last parameter is used to be able to perform the tradeoff between T-count/depth and qubit count (which we discuss in more detail
later on).

In addition to returning the the LUT operation object, the wrapper function also returns the
number of integer and fractional bits required for the input and output. This is required to create the fixed-point registers that the
LUT operation will use as input and output qubit registers. We present the pseudocode of the wrapper function in Algorithm~\ref{alg:makelookup}. We note that we require
the subtraction circuit in Step~\ref{step:subtraction} whenever $x_{\min} \neq 0$ because the LUT is designed in such a way that its list of outputs $\left(f(\hat x_0), f(\hat x_1), \dots\right)$ corresponds 
to inputs in order from $00\dots0$ to $11\dots1$, so the implementation performs the subtraction to have the minimum value $\hat x_0$ be represented by $00\dots00$, the next
value $\hat x_1$ be represented by $00\dots01$ and so on.

\begin{algorithm}[!h]
\caption{The wrapper function \textsc{ApplyFunctionWithLookup} that implements the quantum LUT for arithmetic operations}
\label{alg:makelookup}
\begin{algorithmic}[1]
    \Inputs
        \Statex $f(\cdot)$: Desired arithmetic function to implement
        \Statex $(x_{\min}, x_{\max})$: Input domain (with inclusive bounds)
        \Statex $(\epsilon_{\mathrm{in}}, \epsilon_{\mathrm{out}})$: Maximum allowed error for the input value and precision tolerance for the output value
        \Statex $l$: Number of qubits to be used as swap qubits
    \EndInputs
    \Outputs
        \Statex LUT: Lookup table Q\# operation
        \Statex $(n, p)$: Number of total bits and integer bits respectively required for the input fixed-point register to the LUT
        \Statex $(m, q)$: Number of total bits and integer bits respectively required for the output fixed-point register of the LUT
    \EndOutputs
    \Procedure{ApplyFunctionWithLookup\phantom{abcdefghijklmno} }{$f(\cdot), (x_{\min}, x_{\max}), (\epsilon_{\mathrm{in}}, \epsilon_{\mathrm{out}}), l$}
    \begin{enumerate}
        \item Compute the list of inputs $\hat x_i$ in fixed point representation as well as $(n, p)$ given $ (x_{\min}, x_{\max})$ and $\epsilon_{\mathrm{in}}$
        \item Using the inputs $\hat x_i$, compute the list of outputs $\hat f(\hat x_i)$, the approximation of $f(\hat x_i)$ in fixed-point representation to within a precision of $\epsilon_{\mathrm{out}}$, and $(m, q)$
        \item If $x_{\min} \neq 0$, create a \textsc{Subtraction} circuit that takes in a fixed-point register and subtracts $x_{\min}$ from it \label{step:subtraction}
    \item Create the \textsc{SelectSwap} circuit using $\hat x_i$, $\hat f(\hat x_i)$ and $l$ and append it to the \textsc{Subtraction} circuit
    \end{enumerate}    
    \EndProcedure
\end{algorithmic}
\end{algorithm}

We make use of several Q\# library functions that are convenient to design this wrapper function, such as conversion functions \textsc{BoolArrayAsInt} and 
\textsc{DoubleAsFixedPoint} which easily handle the many transformations from one representation to another which were required throughout the implementation
of the wrapper function. A simple example of how to use the \textsc{ApplyFunctionWithLookup} wrapper function to implement an exponential function can be 
seen in \aref{sec:codeblock}. 

\section{Error analysis of the LUT to implement quantum arithmetic functions}
\label{sec:error}

The final LUT operator guarantees that given an input value $x$ and an arithmetic function $f(\cdot)$, 
it will compute $\hat f(\hat x)$ where $\hat f$ is an approximation of $f$ to within a precision of $\epsilon_{\mathrm{out}}$, and $\hat x$ is an
approximation of the input value $x$ to within the maximum allowed input error tolerance $\epsilon_{\mathrm{in}}$. We note that the total output error is composed of
 $\epsilon_{\mathrm{out}}$ as well as the error propagated by the difference between $\hat x$ and $x$. Therefore if the 
function $f$ is highly oscillatory and/or does not have well-behaved derivatives, then the circuit created to approximate $f(x)$ may 
lead a high output error. However many commonly used arithmetic operations on well defined domains, including everywhere differentiable functions,\footnote{The fact that all everywhere differentiable functions are Lipschitz continuous is a consequence of the 
Mean Value Theorem which was stated and proved in its modern form by Augustin Louis Cauchy in 1823.}
 are Lipschitz continuous i.e. exhibit the property $|f(\hat x) - f(x)| \leq L|\hat x - x|$ for some constant $L$. Therefore for Lipschitz continuous functions, 
the total output error given by the LUT implementation is upper bounded by
\begin{align}
	|\hat f(\hat x) - f(x)| &= |\hat f(\hat x) - f(\hat x) + f(\hat x) - f(x)|  \notag \\
	&\leq |\hat f(\hat x) - f(\hat x)| + |f(\hat x) - f(x)| \notag \\
	&\leq \epsilon_{\mathrm{out}} + L|\hat x - x| \notag \\
	&= \epsilon_{\mathrm{out}} + L\epsilon_{\mathrm{in}}
	\label{eq:total_error}
\end{align} 
In the case of everywhere differentiable functions, $L$ can be replaced with $\sup|\frac{df}{dx}|$. Therefore when implementing the LUT, the user
can easily compute an upper bound of the total output error and use it to determine the desired $\epsilon_{\mathrm{in}}$ and $\epsilon_{\mathrm{out}}$
input values. Being able to define the total output error explicitly is an important feature of the implementation. Firstly, it allows the user to keep track of the error
propagation when using the operator in end-to-end implemenations of quantum algorithms. Secondly, it also makes the LUT implementation a good benchmark for
evaluating the efficiency of single variable bespoke quantum arithmetic circuits.

\section{Comparing the LUT implementations for quantum arithmetic functions to bespoke implementations}
We now compare LUT quantum arithmetic functions with bespoke quantum arithmetic implementations. 
In particular, we look at three examples: the exponential, Gaussian and square root functions.
Given that these arithmetic functions require a large amount of qubits, 
we look at the resources from a fault-tolerant perspective and assume an architecture using Clifford+T-gates as fundamental operations.
In this setting, the resources required to create error-corrected T-gates largely outweigh the resources for creating any Clifford gate~\cite{campbell}, which is
why we focus on the T-count, T-depth and qubit numbers. We note that Q\# has an
in-built resource estimation engine that allows for T-count, T-depth and qubit count calculations for any operation. 
An example of a code snippet used to compute
the resources for a LUT implementation can be found in \aref{sec:codeblock}.

\subsection{Exponential and Gaussian functions}
In \autoref{tbl:expresults}, we compare
the resources for the LUT implementation with the implementation by Poirier for different domains and error tolerances as shown in~\cite{Poirier21}.
In addition to the examples found in~\cite{Poirier21}, we also computed the resources to implement $e^{-x}$ for the domain $[\log(1/2),0]$ 
since in principle any additional power of $e^{-x}$ can be implemented with a bit shift in the fixed point representation of the input by computing the input mod $\log(1/2)$ 
and then computing the argument and bit shifting the results down. The details of how the parameters for \autoref{tbl:expresults} were deduced from~\cite{Poirier21} for the Poirier method
can be found in \aref{sec:poirier}.

\begin{table}[htbp]
\def\tabcolsep{2.5pt}
\caption{Comparing the T-count and Qubit count of exponential and Gaussian functions between the Poirier method~\cite{Poirier21} and the LUT implementations. For LUT implementation, we used 0 swap
qubits in the $e^{-x}$ computations and 5 swap qubits in the $e^{-x^2}$ computations.}
\begin{center}
\begin{tabular}{|c |c|c|c|c|c|c|}
\hline
$f(x)$ & $(x_{\min}, x_{\max})$ & $(\epsilon_{\mathrm{in}}, \epsilon_{\mathrm{out}})$ & \multicolumn{2}{|c|}{T-count} & \multicolumn{2}{|c|}{Qubit Count} \\
\cline{4-7}
 & & & Poirier & LUT & Poirier & LUT \\
\hline
$e^{-x}$ & (0, 10)& $(2^{-3},10^{-7})$ & 6480 & \textbf{1176} & 154 & \textbf{36} \\
\cline{3-7}
  & & $(2^{-4},10^{-9})$ & 17752 & \textbf{2310} & 264 &  \textbf{45}\\
\cline{2-7} 
& (0, 100)& $(1,10^{-7})$ & 3648 & \textbf{1442} & 134 & \textbf{30} \\
\cline{3-7}
  & & $(2^{-1},10^{-9})$ & 11312 & \textbf{2856} & 233 &  \textbf{42} \\
\cline{2-7}
& ($\log(1/2)$, 0)& $(2^{-4},10^{-7})$ & 6480 & \textbf{574} & 154 & \textbf{45} \\
\cline{3-7}
  & & $(2^{-5},10^{-9})$ & 17752 & \textbf{826} & 264 &  \textbf{57} \\ 
\hline
$e^{-x^2}$ & (0, 10)& $(2^{-9},10^{-7})$ & 17872 & \textbf{13560} & \textbf{325} &  3365 \\
\cline{3-7}
  & & $(2^{-11},10^{-9})$ & \textbf{33200} & 36552 & \textbf{465} & 4143 \\
 \cline{2-7}
& (0, 100)& $(2^{-6},10^{-7})$ & \textbf{2816} & 13560 & \textbf{141} & 3331 \\
\cline{3-7}
  & & $(2^{-8},10^{-9})$ & \textbf{12928} & 36552 & \textbf{262} & 4119 \\
\hline
\end{tabular}
\label{tbl:expresults}
\end{center}
\end{table}

We notice that the difference between the resources required to compute $e^{-x}$ and $e^{-x^2}$ are quite large, even though it would seem that 
two can be interconverted in a straightforward way by using a squaring circuit before inputing into $e^{-x}$. However a squaring circuit would add its
own approximation errors which would need to be rigorously taken into account. This would require splitting the error between the two circuits and making up for
any additional errors. Hence the conversion is not so straightforward and could lead to an increase in resources.

\begin{figure}[htbp]
\centerline{\includegraphics[width=0.4\textwidth]{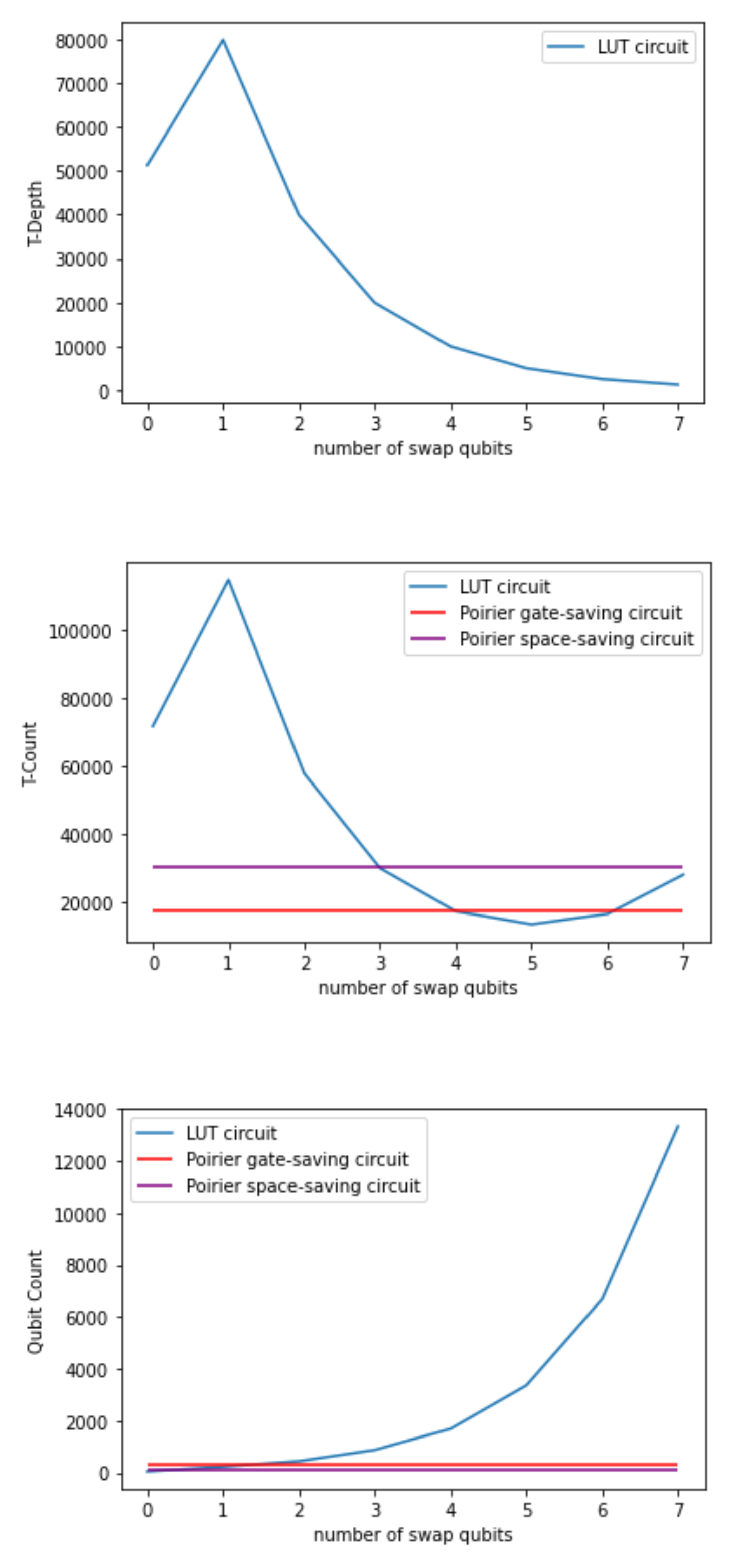}}
\caption{Plots showing resources to implement the LUT for $e^{-x^2}$ with $(x_{\min}, x_{\max})=  (0, 10)$ and $(\epsilon_{\mathrm{in}}, \epsilon_{\mathrm{out}}) = (2^{-9},10^{-7})$.}
\label{fig:tradeoff}
\end{figure}

We would now like to look more closely at the interesting case of the function $e^{-x^2}$ with $(x_{\min}, x_{\max}) = (0, 10)$ and 
$(\epsilon_{\mathrm{in}}, \epsilon_{\mathrm{out}}) = (2^{-9},10^{-7})$. 
We see that there are trade-offs wherein it can be more or less efficient to use the LUT circuit vs the Poirier implementation
depending on what you are trying to optimize. Therefore, we use this example to illustrate the demonstrate the T-gate to qubit count 
trade-offs that can be had when changing the number of swap qubits in the LUT implemention in \autoref{fig:tradeoff}.

For this particular case, we see that the Poirier method is clearly much more efficient in terms of qubit numbers. However the LUT implementation has the ability to trade off for lower T-count/depth
if there are sufficient qubits available.

\subsection{The square root function}
In \autoref{tbl:sqrtresults}, we compare
the resources for the LUT implementation with the ESOP based synthesis~\cite{esop} implementation and the implementation by Dutta et al. for different domains and error tolerances taken from
the results as shown in~\cite{dutta}. The details of how the parameters in \autoref{tbl:sqrtresults} were deduced from~\cite{dutta} ESOP based synthesis and Dutta et al. methods
can be found in \aref{sec:dutta}.

\begin{table*}[htbp]
\caption{Comparing the T-count and Qubit count for the square root function between the ESOP based synthesis and Dutta et al. methods~\cite{dutta} and the LUT implementations. 
For the LUT implementations, in each instance we used the number of swap qubits that optimized for T-count.}
\begin{center}
\begin{tabular}{|c|c|c|c|c|c|c|c|}
\hline
 $(x_{\min}, x_{\max})$ & $(\epsilon_{\mathrm{in}}, \epsilon_{\mathrm{out}})$ & \multicolumn{3}{|c|}{T-count} & \multicolumn{3}{|c|}{Qubit Count} \\
\cline{3-8}
 & & ESOP & Dutta et al. & LUT & ESOP & Dutta et al. & LUT \\
\hline
(0, 4) & $(2^{-5}, 2^{-5})$ & \textbf{803} & 27769 & 1288 & \textbf{16} & 210 & 269 \\
\hline
(0, 8)  & $(2^{-7}, 2^{-7})$ & \textbf{1746} & 53011 & 4216 & \textbf{22} & 288 & 1317\\
\cline{2-8} 
 & $(2^{-9}, 2^{-9})$ & 10242 & 74319 & \textbf{9576} & \textbf{26} & 338 & 3121 \\
\hline
 (0, 16) & $(2^{-9}, 2^{-9})$ & 31075 & 86317 & \textbf{13664} & \textbf{28} & 363 & 3357 \\
\hline
\end{tabular}
\label{tbl:sqrtresults}
\end{center}
\end{table*}

\begin{figure}[htbp]
\centerline{\includegraphics[width=0.4\textwidth]{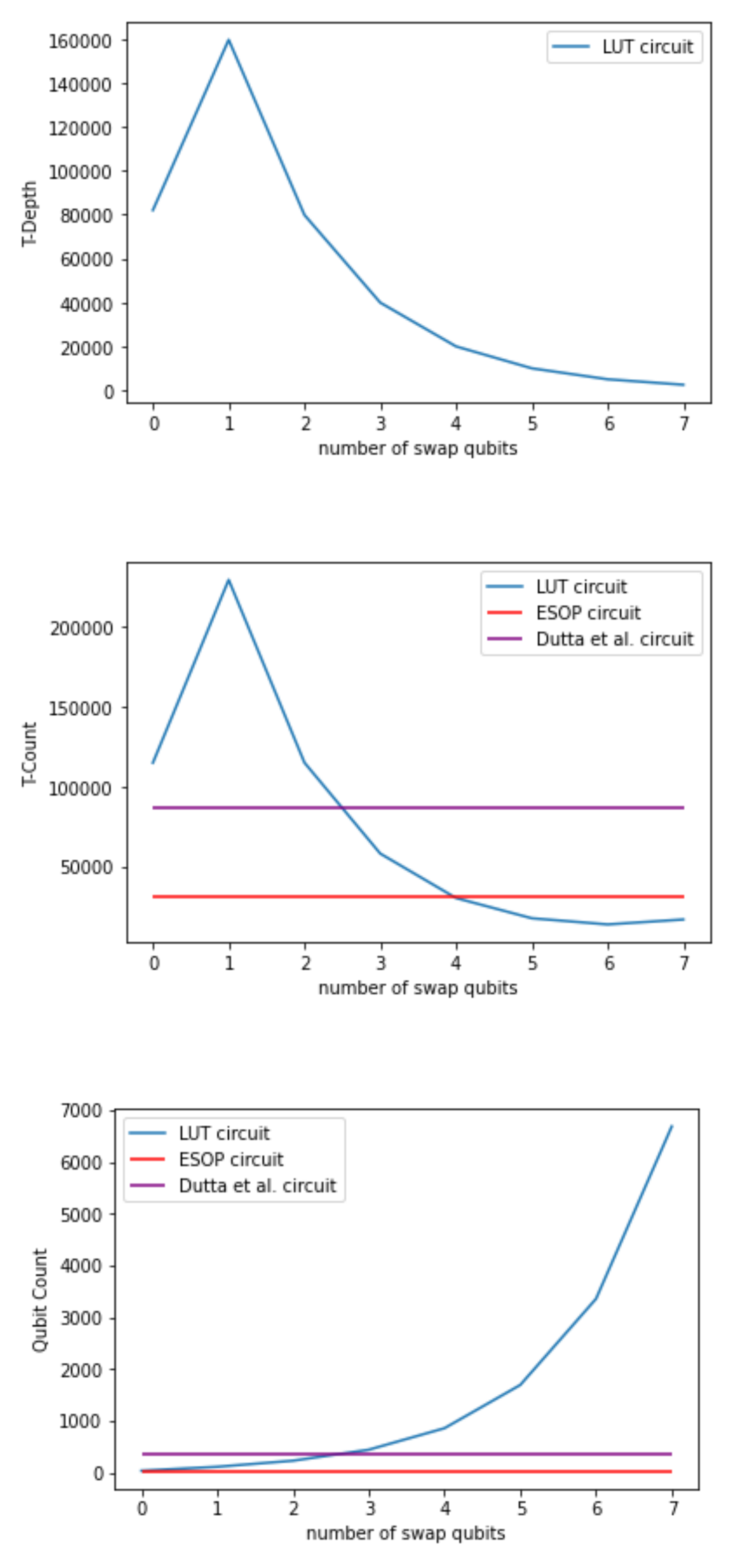}}
\caption{Plots showing resources to implement the LUT for $\sqrt{x}$ with $(x_{\min}, x_{\max})=  (0, 16)$ and $(\epsilon_{\mathrm{in}}, \epsilon_{\mathrm{out}}) = (2^{-9},2^{-9})$.}
\label{fig:sqrttradeoff}
\end{figure}

Similarly to the results for the exponential and Gaussian functions, we can
look at the T-gate to qubit count trade-offs
that can be had when changing the number of swap qubits in
the LUT implemention for an interesting example. In this case, we will focus on
the example with $(x_{\min}, x_{\max}) = (0, 16)$ and 
$(\epsilon_{\mathrm{in}}, \epsilon_{\mathrm{out}}) = (2^{-9},10^{-9})$. The results can be seen in
\autoref{fig:sqrttradeoff}.

Simlarily in this case as well, we notice that the ESOP synthesis based method is clearly much more efficient in terms of qubit numbers, but that LUT implementation has the ability to 
trade off for lower T-count/depth if there are sufficient qubits available. 

\subsection{Remarks and general observations}
In terms of the performance of the LUT implementation with respect to bespoke implementations, we make a few remarks:
\begin{itemize}
	\item The LUT implementation is in general comparable to bespoke implementations of arithmetic functions, where the performance being better or
	worse is dependent on the specific function being implemented. We also find that the LUT function has the ability to trade-off between T-count/depth and
	qubit count such that it can often outperform bespoke implementations in the former but at a large cost of the latter.
	\item By characterizing the parameters $(x_{\min}, x_{\max})$ and $(\epsilon_{\mathrm{in}}, \epsilon_{\mathrm{out}})$ when implementing the function, we were able to make a fair comparison between the LUT implementation 
	and bespoke implementations while also using a benchmark that is relevant to practitioners.
	\item For the simple operations of comparing to a constant, adding or subtracting a constant~\cite{draper}, and multiplying or dividing by a constant,\footnote{For most reasonably sized registers, 
the most efficient way to multiply by a constant is using the ``schoolbook" method and perform a series of addition using any addition circuit e.g.~\cite{draper}. For dividing by a constant, one can simply multiply by the reciprocal.}
	the current state-of-the-art bespoke 
	methods widely outperform the LUT implementation. This can be seen by noting that 
	the T-counts, T-depths and qubit counts of those operations are linear in the number of input qubits with a small constant factor, which is favorable with respect to
	\autoref{eqn:t-depth} and \autoref{eqn:qubit-count}.
\end{itemize}

\section{Conclusion and Discussion}

In this paper, we have presented an implementation of a function that can create a quantum lookup table for any single-variabled
arithmetic functions. The implementation only requires the desired function, input domain, allowed error tolerances and number of swap
qubits as input, which are the parameters that practitioners will typically focus on when implementing
quantum arithmetic operations within a larger end-to-end algorithm. The implementation then takes care of all the appropriate possible
outputs for the LUT and the actual LUT operation itself. It will also output the number or required integer and fractional input and output bits
such that the user can easily create the required fixed point input and output registers used as arguments to the LUT operation. We then discuss how the
approximation error of the operator can be understood. This understanding allows for the implementation to be used in two settings: (1) one can easily use it
 and keep track of its approximation errors when implementating in end-to-end algorithms and (2) it can be used as a well-defined benchmark for evaluating 
the efficiency of single variable bespoke quantum arithmetic circuits. Finally, we
demonstrated how our implementation can be used to create LUT operations for quantum arithmetic functions, and 
compared them to state-of-the-art bespoke quantum arithmetic functions using in-built Q\# functionalities to compute T-count, T-depth and qubits counts.

As a next step, one can think about implementing a function that, given a desired arithmetic operation, input domain and error tolerances, would
automatically compute which between the LUT implementation or the bespoke implementation of said operation is more efficient and return the
appropriate operation object with all the required parameters to create the fixed-point registers. One can also think of incorporating a feature that performs
linear or more complicated interpolations after the quantum arithmetic operation which may allow for overall lower output errors for the same amount of gate and qubit resources. 
One can also think of slightly restructing the LUT implementation
such that instead of the implementation outputting an operation object with the required parameters to create the appropriate fixed-point registers, the function
would encompass an object that would itself handle the creation and propagation of these registers.

Finally, an interesting line of work would be to try and understand more about which general classes of single variable quantum arithmetic functions would we expect LUT implementations
to perform better compared to other (e.g. reversible or QFT) implementations.

\appendices

\section{Q\# code to implement the LUT in a program}
\label{sec:codeblock}
An example of a Q\# file using the \textsc{ ApplyFunctionWithLookupTable} function to implement the LUT function can be found in the code snippet in \autoref{fig:codeblock}.
The example code below was used to implment the function $e^{-x}$ for $(x_{\min}, x_{\max})=  (0, 10)$, 
$(\epsilon_{\mathrm{in}}, \epsilon_{\mathrm{out}}) = (2^{-3},10^{-7})$ and 0 swap qubits.

\begin{figure}
\scriptsize
\begin{Verbatim}[frame=single]
namespace ResourceEstimate {

  open Microsoft.Quantum.Arithmetic;
  open Microsoft.Quantum.Math;
    
  function ExpInv(x: Double) : Double {
    return ExpD(-x);
  }

  @EntryPoint()
  operation Main() : Unit {
    let xmin = 0.0;
    let xmax = 10.0;
    let epsin =  PowD(2.0, -3.0);
    let epsout = 1e-7;
    let numswap = 0;
    let lookup = ApplyFunctionWithLookupTable(
      ExpInv, (xmin, xmax), epsin, epsout, numswap);
    use input = Qubit[lookup::IntegerBitsIn + 
      lookup::FractionalBitsIn];
    use output = Qubit[lookup::IntegerBitsOut + 
      lookup::FractionalBitsOut];
    lookup::Apply(FixedPoint(
      lookup::IntegerBitsIn, input), 
      FixedPoint(lookup::IntegerBitsOut, 
      output));
  }
}
\end{Verbatim}
\caption{Code snippet to implement the LUT function in Q\#}
\label{fig:codeblock}
\end{figure}

Once the file is created, the command to run the resource estimator is \texttt{dotnet run -s ResourcesEstimator}. 
To implement any other single variable arithmetic function (either user-defined or native), the user simply needs to replace \textsc{ExpInv} in the \textsc{Main} operation
with the desired function.

\section{Deducing the parameters for \autoref{tbl:expresults} from~\cite{Poirier21}}
\label{sec:poirier}
The resources displayed in \autoref{tbl:expresults} for the Poirier method were taken from Table II
in~\cite{Poirier21}. The parameters $x_{\min}$, $x_{\max}$, and $\epsilon_{\mathrm{out}}$, as well as the qubit
count for the Poirier method are explicitly stated in~\cite{Poirier21} and were taken directly as is. 
The parameter $\epsilon_{\mathrm{in}}$ was determined
by looking at the number of input qubits $n$ and computing the value represented by the least significant
bit given a register representing $x_{\max}$ assuming the register is in a fixed-point binary representation. 
For example in the case where $0\leq x \leq 10$ with $d=7$ input qubits, we note that the most siignificant bit has to
represent at least $2^3$, so the least significant bit represents $2^{3-7} = 2^{-4}$ which gives us our value for 
$\epsilon_{\mathrm{in}}$ (we have assumed  
no sign bit). And finally to compute the T-counts, we multiply the Toffoli counts given in~\cite{Poirier21} by 4 where we have assumed 
the efficient implementation of the Toffoli gate using Clifford+T gates given in as~\cite{jones}.
For the $(x_{\min}, x_{\max})= (\log(1/2,) 0)$ case, we assumed the same register sizes as for  $(x_{\min}, x_{\max})= (0, 10)$
(i.e. $(n, d, m) = (21, 7, 7)$ for $\epsilon_{\mathrm{out}} = 10^{-7}$ and $(n, d, m) = (32, 8, 8)$ for $\epsilon_{\mathrm{out}} = 10^{-9}$) and computed
the values for $\epsilon_{\mathrm{in}}$ in the same way as for the other other cases.

\section{Deducing the parameters for \autoref{tbl:sqrtresults} from~\cite{dutta}}
\label{sec:dutta}
The resources displayed in \autoref{tbl:sqrtresults} for the ESOP based synthesis and Dutta et al. methods were taken from Fig 5.a
in~\cite{dutta}. All the parameters were taken based on the number of integer and fractional qubits used to represent the input
fixed point representation, given by $m$ and $n$ respectively. It was assumed that the output registers also contained $m$ integer qubits 
and $n$ fractional qubits. So we calculated the domains as $(x_{\min}, x_{\max})= (0, 2^m)$ and the errors as $\epsilon_{\mathrm{in}} = \epsilon_{\mathrm{out}} = 2^{-n}$.
The T-counts and qubit counts were explicitly stated in the table and were used as is in \autoref{tbl:sqrtresults}. The precise value of $x_{\max}$ for those results were 
actually $2^{m-1} + 2^{m-2} +\dots+ 2^{0} + 2^{-1}+\dots+2^{-n}$, but we computed the metrics
for the LUT implementation using  $x_{\max}=2^m$, thus giving slightly more conservative results. 
\section*{Acknowledgment}

The authors would like to thank Dave Clader, Wim van Dam, Vadym
Kliuchnikov, Farrokh Labib, Prakash Murali, and Nikitas Stamatopoulos for their help with the manuscript.

\vspace{12pt}

\end{document}